\documentclass[conference, 10pt]{IEEEtran}
\IEEEoverridecommandlockouts
\usepackage{cite}
\usepackage{amsmath,amssymb,amsfonts}
\usepackage{algorithmic}
\usepackage{graphicx}
\usepackage{subfig}
\usepackage{textcomp}
\usepackage{booktabs}
\usepackage{hyperref}

\begin{document}

\title{Accelerating Channel Estimation and Demodulation of Uplink OFDM symbols for Large Scale Antenna Systems using GPU}

\author{\IEEEauthorblockN{Bhargav Gokalgandhi\IEEEauthorrefmark{1},Christina Segerholm\IEEEauthorrefmark{2},Nilanjan Paul\IEEEauthorrefmark{3},Ivan Seskar\IEEEauthorrefmark{4}}
\IEEEauthorblockA{WINLAB, Rutgers University \\
671, US-1, North Brunswick Township, NJ 08902 \\
Email: \IEEEauthorrefmark{1}bgokal@winlab.rutgers.edu,\IEEEauthorrefmark{2}csegerholm@gmail.com,\IEEEauthorrefmark{3}nilanjan@winlab.rutgers.edu,\IEEEauthorrefmark{4}seskar@winlab.rutgers.edu}
% \and
% \IEEEauthorblockN{2\textsuperscript{nd} }
% \IEEEauthorblockA{\textit{WINLAB} \\
% \textit{Rutgers University} \\
% \textit{North Brunswick, NJ, USA} \\
% \textit{csegerholm@gmail.com}}
% \and
% \IEEEauthorblockN{3\textsuperscript{rd} Nilanjan Paul}
% \IEEEauthorblockA{\textit{WINLAB} \\
% \textit{Rutgers University} \\
% \textit{North Brunswick, NJ, USA} \\
% \textit{nilanjan@winlab.rutgers.edu}}
% \and
% \IEEEauthorblockN{4\textsuperscript{th} Ivan Seskar}
% \IEEEauthorblockA{\textit{WINLAB} \\
% \textit{Rutgers University} \\
% \textit{North Brunswick, NJ, USA} \\
% \textit{seskar@winlab.rutgers.edu}}
}

\IEEEpubid{\makebox[\columnwidth]{978-1-5386-9223-3/19/\$31.00~\copyright2019 IEEE \hfill} \hspace{\columnsep}\makebox[\columnwidth]{ }}

\maketitle

\IEEEpubidadjcol

\begin{abstract}
Increase in the number of antennas in the front-end increases the volume of data to be processed at the back-end. This establishes a need for acceleration in back-end processing. To solve the issue of high volume data processing at back-end, a GPU is utilized. Acceleration for Least Squares channel estimation and demodulation of uplink OFDM symbols is provided by using a combination of CPU and GPU at the back-end. Single user uplink scenario is implemented in near real-time manner using the USRP platform present in the Large scale antenna systems in ORBIT Testbed. The number of antennas and FFT length are varied to provide different scenarios for comparison. The performance of both CPU and GPU is compared for each process.
\end{abstract}


\section{Introduction}\label{sec:intro}

For next-gen wireless systems, Massive MIMO (Multiple Input Multiple Output) systems play an important part in providing diversity and spatial multiplexing for increasing the overall throughput of a wireless system. Massive MIMO systems are used in mmWave communication for providing diversity and increasing the transmit and receive gain of the systems, and can provide spatial multiplexing which can satisfy the high throughput requirements in multi-user systems like LTE and Wi-Fi.

Large Scale Antenna systems, which are used for Massive MIMO implementation, are being included in various standards in next-gen systems, and so flexibility is very important at the front-end for both testing and application of standards, in terms of front-end radio features as well as back-end processing applications for precoding and decoding of data. Software Defined Radios (SDRs) provide flexibility in terms of front-end radio features such as carrier frequency, sampling rate, front-end gain, etc. The flexibility at front-end combined with software based back-end processing can be very useful in applying existing as well as new algorithms for uplink and downlink processing with ease and very less use of additional hardware resources.

\begin{figure}[t!]
	\includegraphics[width=\linewidth]{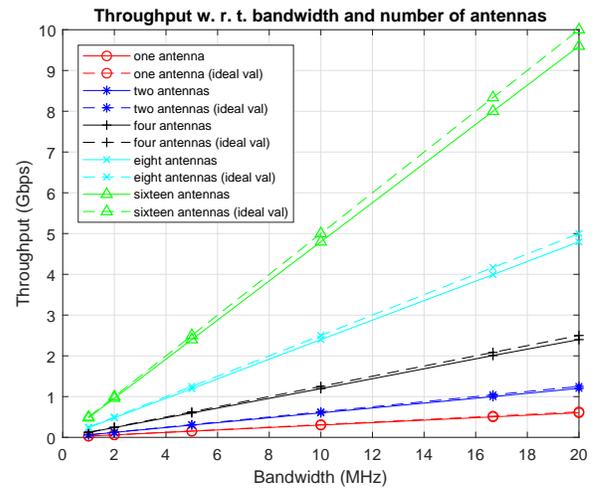}
    \caption{The front-end to back-end throughput with respect to variations in the number of antennas and bandwidth per antenna}
    \label{fig:thr_vs_samp}
\end{figure}

Fig. \ref{fig:thr_vs_samp} shows the maximum throughput, both theoretical and practical values, from front-end SDRs to back-end servers based on the bandwidth and the number of antennas. This is taken from simultaneous data reception with variation in the number of antennas and bandwidth of the Massive MIMO mini-rack present in ORBIT Testbed in WINLAB. Detailed explanation of each part of the front-end system will be in Section \ref{sec:impl}. It is seen that there is a linear increase in throughput with respect to both, increase in the number of antennas and increase in bandwidth, as seen in Fig. \ref{fig:thr_vs_samp}. Data is sent to back-end servers via a 10 Gigabit ethernet link. The 10 Gigabit link entails a maximum limit to the number of antennas and the bandwidth that can be used for wireless transmission and reception.

Considering the increase in data volume as shown in Fig. \ref{fig:thr_vs_samp}, fast and efficient performance of back-end processing systems is needed. Using hardware based solutions such as FPGA and ASIC seems as an attractive idea. But hardware based implementation does not provide enough flexibility for application of different algorithms since the hardware architecture will be tailored for specific algorithms and standards to provide the necessary computing gain. Also hardware based implementation with flexibility can be expensive as compared to its software counterpart. Flexibility being very important for both application and testing of varying algorithms at the back-end, this paper uses software based solutions for application of the algorithms. 

But, implementation of algorithms using software needs speed and efficiency, especially considering the high data volume. A GPU can be used for acceleration of demodulation for real-time processing of received signals GPU can be used to provide the required parallelization which effectively increases speed of computation. Some examples of acceleration for back-end processing of communication systems are shown in \cite{li2015accelerating,li2016decentralized}. To show the acceleration in received signal processing, this paper considers the uplink scenario, where a single user sends OFDM data to a multi-antenna base station. We compare the performance of single core implementation using Intel Xeon CPU \cite{intelcpu} with a Nvidia Tesla K40m GPU \cite{nvidiagpu} for channel estimation and demodulation of OFDM symbols. Performance is compared on a symbol level so that the execution time of each algorithm used can be recorded which in turn shows the effect of each algorithm on the overall execution time. %Using the experimental results of this paper, it will be shown that utilizing GPUs can be advantageous for back-end processing of Large Scale Antenna Systems because of the acceleration provided.

The outline of the paper is as follows. Section \ref{sec:impl} gives details of systems being used with the details and timeline for implementation. Section \ref{sec:exp} shows the results of the experiments and explains the output. Finally, Section \ref{sec:conc} concludes the paper and gives some remarks about the future directions to be taken.
\section{Implementation Details}\label{sec:impl}

\subsection{Components used for implementation}

For implementing the demodulation of OFDM symbols, we use ORBIT Testbed present in WINLAB, Rutgers University. The hardware and software components used are as follows,

\subsubsection{ORBIT Testbed}
\begin{figure}
\includegraphics[width=\linewidth]{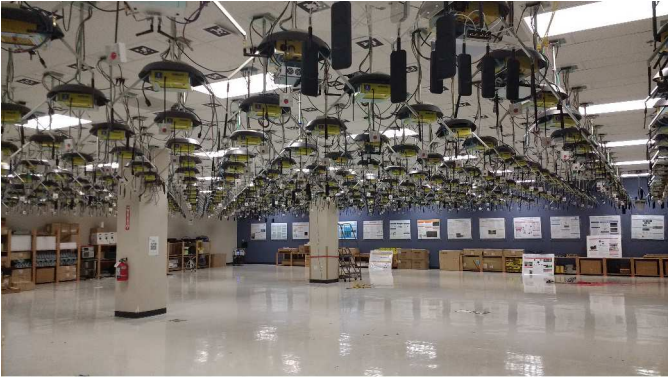}
    \caption{ORBIT Testbed in WINLAB, Rutgers University}
    \label{fig:orbit}
\end{figure}
ORBIT is the largest academic indoor wireless testbed. It consists of a 20x20 grid of computing nodes with more that 100 SDRs connected to the nodes \cite{raychaudhuri2005overview},\cite{orbiturl}. All the nodes are connected to a central console from which control functions can be performed on every node. The testbed also has small sandboxes for smaller experiments and testing codes before conducting actual large-scale experiments. The testbed can be remotely accessed from any part of the globe with an Internet connection. Access to the central console of the testbed is provided from which each node can be controlled and accessed.

\subsubsection{USRP B210 and X310 Software Defined Radio (SDR)}
The SDRs in the grid are various types of USRPs (Universal Software Radio Peripherals) such as N210s, B210s and X310s \cite{usrpx310,usrpb210,ettus2015universal}. USRPs consist of 2 radio front-ends, to which 4 antennas are connected, 2 per front-end giving Full Duplex communication capabilities. Each front-end is controlled by a common motherboard consisting of an FPGA with blocks for accessing the front-end radios and for up-conversion and down-conversion from and to baseband respectively. Various RF parameters can be set for each front-end from the node via the motherboard.

\subsubsection{Massive MIMO Racks}
\begin{figure}
\centering
\includegraphics[width=\linewidth]{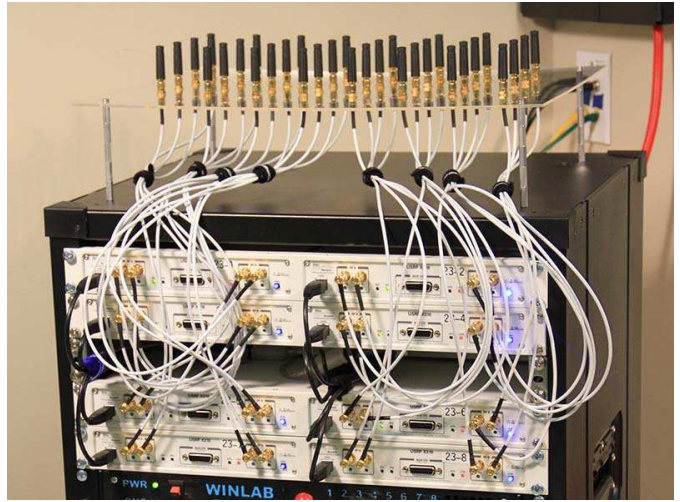}
    \caption{Massive MIMO mini-racks. 4 such racks on each corner of the ORBIT Testbed. Each rack consists of 8 USRP X310s and 32 antennas}
    \label{fig:mimo_rack}
\end{figure}
On the corners of the testbed, there are 4 Massive MIMO mini-racks, as shown in Fig. \ref{fig:mimo_rack}. Each rack has 32 antennas connected to 8 USRP X310s. These X310s are connected to a switch, which connects the USRPs to back-end servers via a 10 Gigabit Ethernet connection. Also, the USRPs are connected to all the nodes in the grid via a 1 Gigabit Ethernet connection. Since there are 4 such mini-racks, there are a total of 128 antennas which can provide co-located as well as semi-distributed sub-6 GHz Massive MIMO.

\subsubsection{OctoClock}
The nodes on the corners of the testbed are synchronized in both time and frequency using an OctoClock \cite{octoclock}. The OctoClock converts synchronizing signals received from GPS satellite to PPS (Pulse Per Second) signal for time synchronization, and 10 MHz clock signal for frequency synchronization. These nodes in the corner of the testbed can be synchronized externally in both time and frequency and can be used to perform collective experiments over multiple nodes.

\subsubsection{Intel Xeon CPU E5-2630 v3}
This CPU has 8 physical cores and 16 logical cores (threads). It has a processor base frequency of 2.4 GHz, 64 GB of memory with maximum memory bandwidth of 59 GBps, and 20 MB of cache for fast memory access. Detailed specifications are given in \cite{intelcpu}.

\subsubsection{Nvidia Tesla K40 GPU}
This GPU has 2880 processor cores with a memory clock of 3 GHz, 12 GB of device global memory, and a constant memory of 64 kB. Kepler architecture based GK110 chip is used in the GPU. This chip has a CUDA compute capability of 3.5 which adds features such as dynamic parallelism and Hyper-Q. More specifications are given in \cite{nvidiagpu,gpuk110}. 

\subsubsection{CUDA}
CUDA is a platform created by Nvidia and used for acceleration of processes using parallel computing. CUDA based drivers are used to run applications on Nvidia GPUs. CUDA functions and APIs (Application Programming Interfaces) can be used to parallelize and speed-up the processes by tailoring the implementation to the Nvidia GPU architectures. %Using CUDA, GPUs can be used to perform general purpose computing in a much easier fashion without the required advanced knowledge of graphics programming.

\subsection{Implementation specifications and outline}

\begin{figure}
\centering
\includegraphics[width=0.8\linewidth]{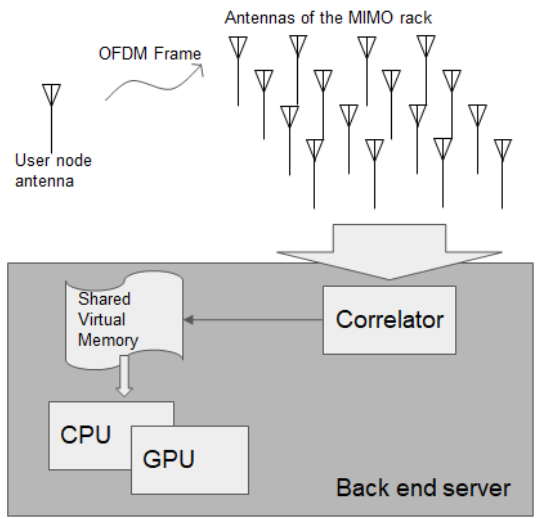}
    \caption{Implementation outline}
    \label{fig:impl_timeline}
\end{figure}

Fig. \ref{fig:impl_timeline} shows the timeline of the implementation. For implementing the uplink scenario and OFDM demodulation, we use a wireless node in the ORBIT Testbed for transmission of OFDM symbols. The OFDM frames are generated in Matlab. The OFDM frames generated consist of a maximum length PN sequence for packet detection. The first OFDM symbol after the PN sequence is the channel estimation symbol. All OFDM symbols after the channel estimation symbol are the QAM modulated data symbols. This OFDM frame is transmitted using the wireless node in the ORBIT grid. The frame is received on all the antennas of the MIMO rack in the testbed. The grid node and the all the USRPs in the MIMO rack are synchronized using OctoClock for time and frequency synchronization. 

The received signal is down-converted to baseband and then transferred to the back-end servers via 10 Gigabit ethernet link. At the back-end servers, the received baseband signal is correlated with the same maximum length PN sequence to find the beginning of the OFDM packet sequence. Using this correlation, the offset to the first OFDM frame is calculated. From there each OFDM frame is saved to a shared memory on a symbol-to-symbol basis for further processing by CPU and GPU applications.

\begin{itemize}
\item \textbf{Shared Virtual Memory:} Shared Virtual Memory is a virtual memory space that can be created by a user level application and accessed by multiple applications via shared memory IDs. In an architecture with multiple applications, a master application can create and free the shared memory with read-write access, while the slave applications only have read-write access to the shared memory. The shared memory has indexes to large arrays used to store the OFDM frames. Shared memory helps facilitate real-time computation by allowing multiple applications simultaneous access to the OFDM frames.
\end{itemize} 

The back-end applications then reads the OFDM symbol from the shared memory in the same manner it is written into the shared memory. The first OFDM symbol read from the shared memory is used for block based Least Squares channel estimation \cite{shen2006channel}. After channel estimation, each OFDM symbol is used for demodulation. The demodulation is performed in a symbol-to-symbol basis so that the performance can be averaged over multiple symbols. Before channel estimation or demodulation, the cyclic prefix for each OFDM symbol is removed and FFT is computed. Computation is performed for the symbol of each received antenna. All demodulated symbols are then aggregated and stored in a file for further offline processing and visualization. 

\subsection{GPU implementation details}

\begin{figure}
\centering
\includegraphics[width=0.8\linewidth]{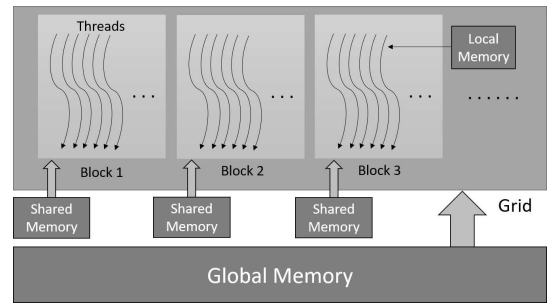}
    \caption{GPU architecture \cite{gpuk110}}
    \label{fig:gpu_arch}
\end{figure}

\begin{figure}
\centering
\includegraphics[width=0.8\linewidth]{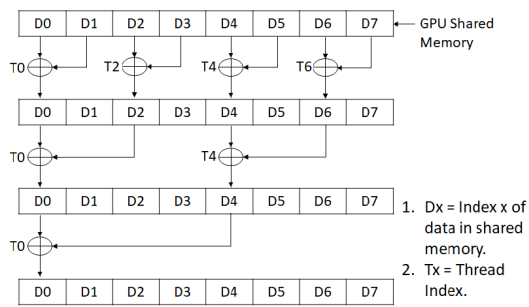}
    \caption{Example of Array Reduction technique used for parallelizing summation \cite{reductiongpu}}
    \label{fig:array_reduction}
\end{figure}

For implementation, the GPU architecture show in Fig. \ref{fig:gpu_arch} is taken into consideration. The GPU processors are divided in a grid. Each thread has a Local Memory, each block consists of a Shared Memory which all threads within the block can access and all blocks have an access of Global Memory \cite{gpuk110}.

Consider a multi-antenna system with $N$ number of antennas receiving OFDM symbols with $M$ sub-carriers. For channel estimation and demodulation, processes such as reading of OFDM symbols, dropping of Cyclic Prefix and FFT are common. For FFT, the batch based FFT functions are used to accelerate the processing and FFTs for symbols of all antennas are computed in parallel. For channel estimation, Least Squares algorithm, which is division of each received sub-carrier of the first received OFDM symbol of each received antenna with the Pilot symbols \cite{shen2006channel}. $M$ threads, each consisting a single sub-carrier value in their local memory and $N$ blocks, each for one receiving antenna, consisting of $M$ threads are used for parallelization of Least Squares algorithm. 

For combining the received OFDM symbols of all antennas, Maximal Ratio Combining (MRC) \cite{goldsmith2005wireless} is used. $M$ blocks are used in parallel in a single grid each consisting of $N$ threads. Each block has data of a single sub-carrier of the OFDM symbols of all received antennas stored in shared memory. Then, using parallel reduction, the summation of all values of data stored in shared memory is performed \cite{reductiongpu}. An examples of parallel array reduction for summation is shown in Fig. \ref{fig:array_reduction}. Parallelizing summation reduces the computational complexity to $O(logN)$ as compared to sequential summation for which the complexity is $O(N)$. %As shown in Fig. \ref{fig:array_reduction}, the first step will use $N/2$ threads for addition of $N$ data values with each thread adding values from adjacent data index. Then, the stride between data index is doubled and the number of threads are halved for each proceeding step till only $1$ thread is required for final summation.

Processes such as Cyclic Prefix dropping and \textit{fftshift} are implemented in CPU as these require data transfer within memory for which the processing time is less in CPU.
\section{Experiments and Results}\label{sec:exp}

\begin{table}
\begin{center}
  \begin{tabular}{c|c}
    \toprule
    \textbf{Parameter} & \textbf{Value} \\
    \hline
    Number of transmitting antennas & 1 \\
    Number of receiving antennas & 1 to 16 \\
    Carrier Frequency & 5.4 GHz \\
    Bandwidth per antenna & 10 MHz \\
    Modulation & OFDM \\
    Channel Estimation & Least Squares \\
    Sub-carriers & 64, 1024 \\
    Cyclic Prefix length & 16, 72 resp. \\
    %Constellation & BPSK, QPSK,\\
    %& 16QAM, 64QAM,\\ 
    %& 256QAM \\
    Number of modulated & 100000 approx. \\
    samples \\
    PN Sequence length & 255 \\
    \bottomrule
  \end{tabular}
\end{center}
\caption{Parameters for the experiments}
\label{tab:exp_param}
\end{table}

The outline mentioned in Section \ref{sec:impl} is used to compare the performance of the CPU and GPU for OFDM demodulation. For comparison of performance, number of antennas and FFT size are varied. As shown in Table \ref{tab:exp_param}, the FFT lengths of OFDM symbols used are 64 and 1024. For each of the two FFT lengths, the antennas are varies from 1 to 16. $5$ GHz frequency band is used with 10 MHz bandwidth. While Fig. \ref{fig:thr_vs_samp} shows that the maximum limit of network is reached at $20$ MHz, data volume for bandwidth greater than $10$ MHz cannot be handled by the server systems. The number of symbols in the OFDM frame vary according to the FFT length and the number of QAM samples to be modulated in an OFDM frame. The execution time for each function be performed is recorded. Acceleration is then calculated using the CPU and GPU times for each function. The demodulation time of each OFDM symbol is averaged and compared.

% The acceleration for channel estimation is shown in Fig. \ref{fig:speed_up_chan_est}. The calculation of the time taken for channel estimation consists of reading of OFDM symbols from shared memory, removing the cyclic prefix, FFT, Least Squares channel estimation, taking conjugate of channel estimates and the summation of all elements of the channel vector. Channel conjugate and calculation of summation of all elements of channel vector during channel estimation is important as it is used for MRC for all symbols and helps with acceleration as it is calculated beforehand.

\begin{figure}[t!]
\includegraphics[width=\linewidth]{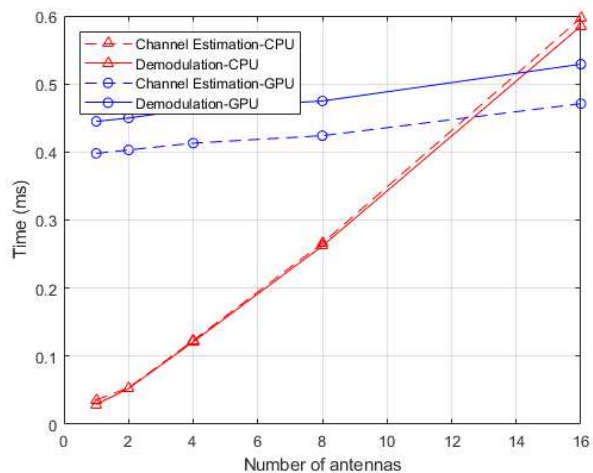}
    \caption{Execution time for channel estimation and demodulation w.r.t. number of antennas for OFDM symbols with 64 sub-carriers}
    \label{fig:time_64fft}
\end{figure}

\begin{figure}[t!]
\includegraphics[width=\linewidth]{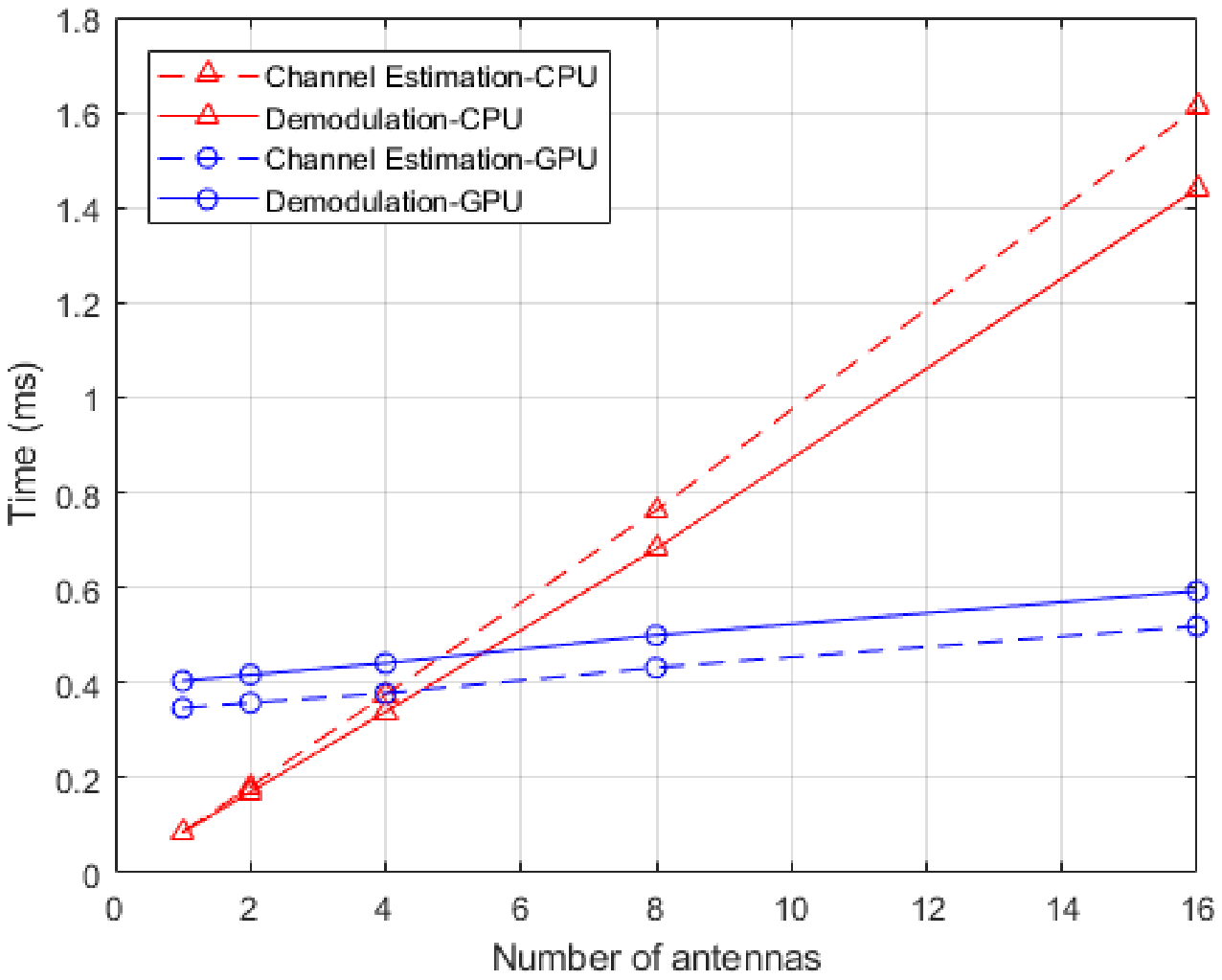}
    \caption{Execution time for channel estimation and demodulation w.r.t. number of antennas for OFDM symbols with 1024 sub-carriers}
    \label{fig:time_1024fft}
\end{figure}

\begin{figure}[t]
\includegraphics[width=\linewidth]{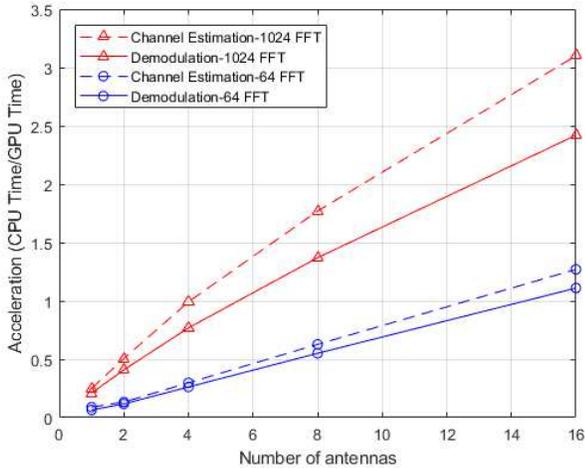}
    \caption{Average acceleration for Least Squares channel estimation and demodulation w.r.t. number of antennas. Graphs are for 1024 length FFT and 64 length FFT based OFDM symbols}
    \label{fig:speed_up}
\end{figure}

Figs. \ref{fig:time_64fft} and \ref{fig:time_1024fft} show the the execution times for CPU and GPU for Least Squares channel estimation and demodulation of OFDM symbols with $64$ and $1024$ sub-carriers respectively. The execution time consists of symbol read, dropping of Cyclic Prefix, FFT, and Least Squares for channel estimation or MRC for demodulation. The increase in execution time for CPU is drastic whereas for GPU is minimal due to sub-carrier and antenna level parallelization. When the number of antennas is low, the CPU execution time is less than GPU. Increase in execution time for GPU is due to data transfer latency between CPU and GPU. Since the acceleration provided due to parallelization is not enough to offset the data transfer latency, the time consumption when using GPU is much higher than that of CPU.

Fig. \ref{fig:speed_up} shows the acceleration obtained after using a GPU. Due to the acceleration provided by FFT computation and computation of Least Squares for OFDM, for 1024 sub-carrier based OFDM symbols, average speed-up of channel estimation is much better than for 64 sub-carrier based OFDM symbols. This dependence on the FFT length creates a limit on the usability of the back-end server systems. Acceleration on a symbol-to-symbol basis can be provided only for very large scale antenna systems where the functions for which the high number of GPU cores can be utilized simultaneously.

%To decrease this dependence on the FFT length and number of antennas, a frame-by-frame basis of computation must also be considered. By parallelizing  the computation of multiple symbols within the OFDM frame, acceleration on a much larger scale can be obtained.

% \begin{figure}[t]
% \includegraphics[width=\linewidth]{ser_and_thput_num_ants_small.eps}
%     \caption{Sample Error Rate and Throughput vs Number of antennas obtained for various QAM constellations}
%     \label{fig:ser_thput}
% \end{figure}

%We also have the Symbol Error Rate (SER) and Throughput analysis for the experiments done as shown in Fig. \ref{fig:ser_thput}. The SER is for QAM symbols and there is no error for BPSK and QPSK constellations due to the diversity gains. This gives maximum possible throughput for QPSK and BPSK for the user. While the error for 64QAM and 256QAM is much higher than 16QAM, considering the bit-per-sample, we get the highest throughput of $50$ Mbps. The error rate does not change much as the number of antennas increase from $8$. This shows the limit of diversity gain provided by the receiver.

\section{Conclusion and Future Work}\label{sec:conc}

To solve the issue of increasing volumes of data to be processed due to increase in the number of antennas, we used GPU as a processing alternative at the back-end to CPU and FPGA. The trend of increase in acceleration with increase in number of antennas shows promise for utilization of GPU for applications like Massive MIMO.

But the increase in speed being dependent on characteristics such as FFT length and number of antennas limits the advantages that GPU based processing can offer. So, we will be considering parallelization on a larger scale, such as that of an OFDM frame, instead of an OFDM symbol.

At this point the bandwidth per antenna is kept constant while the number of antennas are increased. So, if the bandwidth and number of antennas are increased simultaneously, the volume of incoming data cannot be served by a single link or a single server. So, utilization of distributed algorithms and distributed server systems needs to be considered.

The source code can be found at \url{https://github.com/bhargav0410/gpu-accel-ofdm-ls-mrc.git}

%\nocite{*}
%\printbibliography
\bibliographystyle{unsrt}
\bibliography{ref.bib}

\end{document}